\documentstyle[tighten,preprint,eqsecnum,aps,prd,epsf]{revtex}

\def\be{\begin{equation}}
\def\ee{\end{equation}}
\def\tR{\tilde{R}}
\def\tx{\tilde{\bf x}_0}
\def\sa{\sqrt{\alpha'}}

\begin{document}
\draft

\preprint{\vbox{\baselineskip=12pt
\rightline{gr-qc/9708018}
\rightline{}
\rightline{IISc-CTS-5/97}
\rightline{}
}}
\title{Absence of initial singularities in superstring cosmology
\footnote{Based on talk given at Conference on
Big Bang and Alternative Cosmologies: A Critical Appraisal,
Bangalore, India, January 1997.
}}
\author{Sanjay Jain\footnote{Also at Jawaharlal Nehru Centre for Advanced
Scientific Research, Bangalore 560064, and Associate Member of ICTP, Trieste.
Electronic address: {\em jain@cts.iisc.ernet.in}}}

\address{Centre for Theoretical Studies \\ Indian Institute of Science \\
Bangalore 560012, India}

\maketitle
\begin{abstract}

In a universe whose elementary constituents are point particles
there does not seem to be any obvious mechanism for avoiding the
initial singularities in physical quantities in the standard
model of cosmology. In contrast in
string theory these singularities can be absent
even at the level where spacetime is treated classically. This is 
a consequence of the basic degrees of freedom of strings in compact spaces,
which necessitate a reinterpretation of what one means by a very
small universe. We discuss the basic degrees of freedom of a string
at the classical and quantum level, the minimum size of strings
(string uncertainty principle), the t-duality symmetry, and string
thermodynamics at high energy densities, and then describe how these
considerations suggest a resolution of the initial singularity problem.
An effort has been made to keep this writeup self-contained and accessible to
non-string theorists.

\end{abstract}

\narrowtext

\vfil
\pagebreak

\section{Introduction}
\label{sec:intro}
The first and most radical departure of string theory from a theory
of elementary point particles is in the nature of its elementary degrees of
freedom. The rest of the structure of string theory rather conservatively
follows established principles: relativistic invariance
(generalized to include supersymmetry), quantum
mechanics, locality of interaction, and internal mathematical consistency.
Together, these result in new symmetries and properties that
open up conceptual possibilities inconceivable in elementary point
particle theories.
In this note we focus on a few such properties that make it possible
to imagine a resolution of the initial
singularity problem in cosmology. As will be evident, this possibility
is directly traceable
to the basic degrees of freedom of a string.

In the next section we begin with a detailed discussion of the
degrees of freedom in string theory, contrasting them with
elementary point particle theory, and describing the nature of
`particles' in string theory. The `string uncertainty principle'
(which modifies the Heisenberg uncertainty principle to state that
there is a minimum observable size in a world whose fundamental
constituents are strings) is introduced
in terms of string wavefunctions. The `t-duality' symmetry
(indistinguishability of very large and very small universes)
is described. Section III discusses some features of the thermodynamics
of an ideal gas of strings at very high energy densities. This suggests
the possibility that temperature and pressure will have finite
limiting values in a string universe. In section IV the preceding
material is applied to string cosmology. We discuss a thought
experiment for measuring the size of
a very small universe in the context of string theory. This
prompts a shift of perspective regarding what one means by a small universe,
and from the new vantage point
the initial singularity problem disappears (Brandenberger and Vafa 1989).
We conclude with some
cautionary remarks and some questions of possible relevance to
observations. While much of the material presented here is a review
of existing literature, there are some points which do not seem
to have appeared elsewhere. One is
the expression and physical interpretation of the pressure
of an ideal string gas. Another is the derivation of the string uncertainty
principle from string wavefunctions.

\section{The basic degrees of freedom of a string}
\label{sec:dof}
\noindent {\bf Classical string modes in flat euclidean space}:
The classical configuration of a scalar particle is described by
specifying its position, a point in space, say ${\bf R}^d$.
The classical configuration of a
closed bosonic string is similarly described by specifying its position,
in this case a closed curve embedded in space.
The latter can be described by specifying
a continuous map ${\bf x}$ from the unit circle into space,
${\bf x}:S^1 \rightarrow {\bf R}^d$.
If $\sigma \in [0,\pi]$ is a coordinate on $S^1$ and
$x^i \in (-\infty,\infty), \ i=1,\ldots,d$ are coordinates of ${\bf R}^d$,
the map ${\bf x}$ is specified by $d$ continuous functions
$x^i(\sigma)$ that are periodic, $x^i(0)=x^i(\pi)$. The point
${\bf x}(\sigma)=(x^1(\sigma),\ldots,x^d(\sigma))$
traces out a closed curve in space as $\sigma$ traverses
from $0$ to $\pi$.

It is convenient to Fourier decompose the functions $x^i(\sigma)$:
\begin{equation}
\label{fourier}
x^i(\sigma) = x^i_0 + \sum_{n=1}^\infty(x^i_n \cos 2n\sigma +
\tilde{x}^i_n \sin 2n\sigma).
\end{equation}
The infinite-tuple
$({\bf x}_0,\{{\bf x}_n,\tilde{\bf x}_n\})$
equivalently describes the map ${\bf x}$ or the classical string
configuration. ${\bf x}_0 = (1/{\pi})\int_0^\pi d\sigma {\bf x}(\sigma)$
is a `centre of mass' coordinate describing the average position
of the string. ${\bf x}_n,\tilde{\bf x}_n$ describe the extension
of the string in space. It is evident that a string has an infinitely
richer repertoire of classical configurations than a point particle.
E.g., in the configuration where all ${\bf x}_n,\tilde{\bf x}_n$ are zero
the string has no extension and
becomes a point particle at the point ${\bf x}_0$ in space.
If $x^1_1=\tilde{x}^2_1=r$ and all other $x^i_n,\tilde{x}^i_n$ are
zero, the string configuration is a circle of radius $r$ in a plane
parallel to the $x^1x^2$ plane with centre ${\bf x}_0$. If some of
the $x^i_n,\tilde{x}^i_n$ with higher $n$ are also non-zero, the
configuration will become a `wiggly' circle, and wiggles become
finer with increasing $n$ (typical radius of curvature of a wiggle
caused by the $n^{th}$ mode is $\sim r/n$).
The length of the string in a classical configuration specified by
a map ${\bf x}$ is given by $l = \int_0^\pi d\sigma |d{\bf x}/d\sigma|$,
and it is evident that this can be expressed in terms of the
${\bf x}_n,\tilde{\bf x}_n$.

\vskip 0.3cm
\noindent{\bf String wavefunctions; particle like states}:
In the quantum theory a state can be described by a wavefunction, a
complex valued function over the configuration space. For a point
particle the configuration space is just ${\bf R}^d$ (real space),
hence a wavefunction is a map $\psi :{\bf R}^d \rightarrow {\bf C}$.
For a string let us take the configuration space to be the space
$Q$ of infinite-tuples $({\bf x}_0,\{{\bf x}_n,\tilde{\bf x}_n\})$
or, equivalently, the set of maps ${\bf x}:s^1 \rightarrow {\bf R}^d$.
Then a string wavefunction is a map $\psi :Q \rightarrow {\bf C}$,
assigning, to every string configuration ${\bf x}$ or infinite-tuple
$({\bf x}_0,\{{\bf x}_n,\tilde{\bf x}_n\})$, the complex number
$\psi[{\bf x}]\equiv \psi({\bf x}_0,\{{\bf x}_n,\tilde{\bf x}_n\})$.
(We ignore for the moment the subtlety that strictly $\psi$ ought to
be a map from $Q$ modulo diffeomorphisms of the circle.)
Just as for a point particle $\psi^*({\bf x})\psi({\bf x})$ stands
for the probability density of finding the particle at the point
${\bf x}$ in space when it is in the state $\psi$, so similarly
$\psi^*({\bf x}_0,\{{\bf x}_n,\tilde{\bf x}_n\})
\psi({\bf x}_0,\{{\bf x}_n,\tilde{\bf x}_n\})$ represents the
probability density (in the infinite dimensional space $Q$) of finding
the string in the configuration $({\bf x}_0,\{{\bf x}_n,\tilde{\bf x}_n\})$
when it is in the state $\psi$.

It is instructive to consider a few sample string wavefunctions to see
what the string would look like in the corresponding states.

\vskip 0.1cm \noindent
{\bf 1}. $\psi[{\bf x}] = \delta^d({\bf x}_0-{\bf X}_0) \prod_{n=1}^{\infty}
\delta^d({\bf x}_n-{\bf X}_n)\delta^d(\tilde{\bf x}_n-\tilde{\bf X}_n)$.
\hfill\break
This wavefunction has support only over a single classical configuration
described by the infinite-tuple $({\bf X}_0,\{{\bf X}_n,\tilde{\bf X}_n\})$.
Thus an `observation' of the components of the infinite-tuple
$({\bf x}_0,\{{\bf x}_n,\tilde{\bf x}_n\})$ in this state
(assuming such an experiment can be devised, a point we return
to later) would yield a definite answer for each component, with
the conclusion that in this quantum state the string looks like
a classical string at the configuration
$({\bf X}_0,\{{\bf X}_n,\tilde{\bf X}_n\})$.
In particular if the ${\bf X}_n,\tilde{\bf X}_n$ are all zero, the
string would appear to be a point particle (with no extension) localized
at ${\bf X}_0$.

\vskip 0.1cm \noindent
{\bf 2}. $\psi[{\bf x}] = \delta^d({\bf x_0}-{\bf X}_0)
[\prod_{i=1}^d \phi_i(x^i_1)\tilde{\phi}_i(\tilde{x}^i_1)]
[\prod_{n=2}^{\infty}\delta^d({\bf x}_n)\delta^d(\tilde{\bf x}_n)]$.
\hfill\break
Case i) $\phi_i = \tilde{\phi}_i = \phi$ for all $i$ and $\phi$ is a function
that has support only in a small region around the origin (e.g.,
$\phi(x)=e^{-x^2/a^2}$). In such a state, the fourier components
for $n=1$ have a spread of order $a$, and consequently the string
will appear to have an average position still ${\bf X}_0$ but will
now have an indefinite extension in space of order $a$. If $a$
is small compared to the probes, it will still appear particle-like
(a fuzz of size $a$ around ${\bf X}_0$). If $\prod_{i=1}^d \phi(x_1^i)
= \phi(|{\bf x}_1|)$ (as in the example), the fuzz will appear
spherically symmetric.\hfill\break
Case ii) The $\phi_i$ still have support in a small region of order
$a$ but are now different for different $i$. Then the
`fuzz of size $a$ around ${\bf X}_0$' will no
longer appear isotropic. (The spin of the photon, graviton, etc.,
in string theory is due to such internal spatial structure of the
corresponding states. There $a$ is of order Planck length.)
It should also be evident that the visual picture
is not qualitatively very different if some of the higher modes
are also given a fuzz around the origin. Since these states
are not eigenstates of the ${\bf x}_n,\tilde{\bf x}_n$, they are also
not eigenstates of the length operator. When higher modes are allowed
to be nonzero, the expectation value of $l$ increases (more
wiggles in the classical configurations over which the wavefunction
has support).

\vskip 0.1cm \noindent
{\bf 3}. $\psi[{\bf x}] = e^{-i{\bf p.x_0}}
[\prod_{i=1}^d \phi_i(x^i_1)\tilde{\phi}_i(\tilde{x}^i_1)]
[\prod_{n=2}^{\infty}\delta^d({\bf x}_n)\delta^d(\tilde{\bf x}_n)]$.
\hfill\break
If $\phi_i$ are as in 2 this wavefunction represents the same particle-like
string state, but now with a definite centre of mass momentum ${\bf p}$ rather
than a definite centre of mass position ${\bf X}_0$.

The above discussion made no reference to string dynamics. We only
discussed the degrees of freedom in string theory and how particle
like states can be imagined out of strings. It turns out that the
most natural dynamics for strings in fact makes such states appear
as eigenstates of the energy. The latter turn out to be plane waves
for the centre of mass
coordinate and wavefunctions localized around the origin (in fact
harmonic oscillator wavefunctions) for the wiggle modes.

\vskip 0.3cm
\noindent {\bf Dynamics:}
As a single string moves, its trajectory traces out a two dimensional
surface embedded in space. This can be described by introducing a
fictitious parameter $\tau$ taking values in the interval
${\cal{T}} = [\tau_1,\tau_2]$ of the real line
and letting ${\bf x}$ be a map from ${\cal{T}} \times S^1$
into space. The image of ${\cal{T}} \times S^1$ under the map is a string
trajectory. To specify the dynamics for a single string we need
to introduce an action for every such trajectory. It is rather
cumbersome to introduce a relativistic invariant action in terms
of the space variables ${\bf x(\tau,\sigma})$ alone. It is more convenient to
let the time coordinate $x^0$ also be a function of $\tau$ and
$\sigma$ and to write an action for $x(\tau,\sigma) \equiv
(x^0(\tau,\sigma),x^1(\tau,\sigma),\ldots,x^d(\tau,\sigma))
\equiv (x^{\mu}(\tau,\sigma)), \ \mu =0,1,\ldots,d$, which
describes a string trajectory, or `worldsheet', in $d+1$ dimensional
Minkowski spacetime. The action is taken to be proportional to
the area of the worldsheet (in analogy with the
action of a relativistic point particle's trajectory, which is
proportional to the length of the corresponding worldline).
\begin{equation}
\label{action1}
S[x]=-{1 \over {2\pi\alpha'}}\times {\rm area\ of\ worldsheet}
=- {1 \over {2\pi\alpha'}}\int_{\tau_1}^{\tau_2} d\tau\int_0^\pi d\sigma
\sqrt{-\det \gamma},
\end{equation}
where $\gamma_{\alpha \beta}\equiv \partial_{\alpha}x^\mu
\partial_{\beta}x^\nu \eta_{\mu \nu}$ is the metric on the
worldsheet induced from the spacetime metric $\eta_{\mu\nu}
= {\rm diag}(1,-1,\ldots,-1)$, and $\alpha,\beta$ refer to
worldsheet coordinates $\xi \equiv (\xi^\alpha) \equiv (\xi^0,\xi^1)
\equiv (\tau,\sigma)$.

This dynamics has relativistic invariance in that one can define
the generators of spacetime translations $P^\mu$, and of rotations
and boosts $M^{\mu\nu}$ (see Scherk (1975) for a review)
whose Poisson brackets satisfy the Poincare
algebra. The $P^\mu$ are canonically conjugate to the centre of
mass coordinates $x^\mu_0$. The proportionality constant
${1 \over {2\pi\alpha'}} \equiv T$ is called the string tension
or mass per unit length
since the energy $P^0$ of any static classical configuration
$(x^0(\tau,\sigma) = \tau, x^i(\tau,\sigma) = x^i(\sigma))$
turns out to be $T l$, where
$l$ is the length.

The action (\ref{action1})\ is nonlinear in the derivatives
$\partial_\alpha x^\mu$. But it has an infinite local symmetry
corresponding to the reparametrizations $\xi \rightarrow \xi' =
\xi'(\xi)$ which can be used to bring $\gamma_{\alpha\beta}$
into the form $\gamma_{\alpha\beta}(\xi)=\eta_{\alpha\beta}\rho(\xi)$,
with $\eta_{\alpha\beta}={\rm diag}(1,-1)$. In this `conformal gauge'
$\sqrt{-\det \gamma}=\rho=(1/2)\gamma_{\alpha\beta}\eta^{\alpha\beta}$
and the action reduces to the free scalar field form $S = -(1/4\pi\alpha')
\int \ d\tau d\sigma[(\partial_\tau x)^2-(\partial_\sigma x)^2]$.
Substituting the mode expansion
$x^\mu(\tau,\sigma)=x^\mu_0(\tau)+\sum_{n=1}^\infty
[x^\mu_n(\tau)\cos 2n\sigma + \tilde{x}^\mu_n (\tau)\sin 2n\sigma]$
in this action yields $S = \int \ d\tau L$ with
\begin{equation}
\label{lagrangian}
L = -{1 \over {4\alpha'}}\dot{x}_0^2 - {1 \over {8\alpha'}}\sum_{n=1}^\infty
[({\dot{x}}_n^2 - 4n^2 x_n^2) + ({\dot{\tilde{x}}}_n^2
-4n^2{\tilde{x}}_n^2)],
\end{equation}
where the dot denotes derivative w.r.t. $\tau$. Thus the area law
dynamics automatically prescribes a free particle like role for
the centre of mass mode and a simple harmonic oscillator like
role for the wiggle modes $x_n,\tilde{x}_n$ with frequency $2n$.

It is then evident that the quantum states $\psi$ of the system in question
(a single free string in Minkowski spacetime) will be specified
by the set of quantum numbers $(p,\{N_n,\tilde{N}_n\})$ where $p$ is
a momentum conjugate to the centre of mass
mode (and is the eigenvalue of the spacetime translation generator $P$)
and $N_n,\tilde{N}_n$ are harmonic oscillator excitation level
quantum numbers for the modes $x_n,\tilde{x}_n$.

The conformal gauge does not fix the freedom of reparametrizations
completely. All the $x^\mu$ are not independent variables. One
can show that the independent variables can be taken to be
$({\bf x}_0,\{x^I_n,\tilde{x}^I_n\})$ where $I$ goes only over
the `transverse' spatial indices $I=1,\ldots,d-1$. Classically,
once these are known as functions of $\tau$, all others,
$(\{x^d_n,\tilde{x}^d_n\})$ and $(x^0,\{x^0_n,\tilde{x}^0_n\})$,
are determined as functions of $\tau$ by the constraints and hence the string
worldsheet is determined. (Roughly speaking, diffeos of $(\tau,\sigma)$
eat up two spacetime coordinates $x^0,x^d$ excepting the zero mode of $x^d$.)

\vskip 0.3cm
\noindent {\bf The spectrum}:
Quantum mechanically, this means
that string states are characterized by the set of
quantum numbers $({\bf p},\{N_n^I,\tilde{N}_n^I\})$, the
other quantum numbers being determined in terms of them. In particular
the quantum number $p^0 \equiv \epsilon$, eigenvalue of the energy
$P^0$, is given by
\begin{equation}
\label{spectrum1}
\epsilon^2={\bf p}^2 + (2/\alpha')[-({{d-1} \over {12}}) +
\sum_{I=1}^{d-1}\sum_{n=1}^\infty
n(N_n^I + \tilde{N}_n^I)].
\end{equation}
The closure of the quantum Lorentz algebra fixes $d=25$.
This defines the spectrum of the free closed string in ${\bf R}^d$.

The wavefunction of this state in the basis of independent coordinate variables
then follows from inspection of (\ref{lagrangian}):
\begin{equation}
\label{wave1}
\psi_{({\bf p},\{N_n^I,\tilde{N}_n^I\})}({\bf x}_0,\{x_n^I,\tilde{x}_n^I\}) =
e^{-i{\bf p \cdot}{\bf x}_0} \prod_{I=1}^{d-1}\prod_{n=1}^\infty \ \
H_{N_n^I}(\sqrt{n \over {2\alpha'}} x_n^I)
H_{\tilde{N}_n^I}(\sqrt{n \over {2\alpha'}} \tilde{x}_n^I)
e^{-{n \over {4\alpha'}}[(x_n^I)^2 + (\tilde{x}_n^I)^2]},
\end{equation}
where $H_m(x)$ is the $m^{th}$ Hermite polynomial.
\footnote{There is an additional constraint $f(\{N_n^I,\tilde{N}_n^I\})=0$
on the oscillators coming from the fact that there is no preferred
point in the $\sigma$ direction along the string. The form of $f$ is
more complicated than the usual $f=\sum n(N_n^I - \tilde{N}_n^I)$
because the $x_n,\tilde{x}_n$ defined here do not represent the left
and right moving modes respectively. We henceforth assume that
$({\bf p},\{N_n^I,\tilde{N}_n^I\})$ in (\ref{spectrum1})\ and
(\ref{wave1}) are such that this constraint is satisfied.}

It is interesting to compare this with the third wavefunction discussed
earlier. The $\phi_i(x_1)$ there are replaced by harmonic oscillator
wavefunctions whose width is order $\sqrt{\alpha'}$. The $\delta$-functions
of the higher $n$ modes are also replaced by harmonic oscillator wavefunctions
of width $\sim \sqrt{\alpha'/n}$. Thus a string in a state with quantum numbers
$({\bf p},\{N_n^I,\tilde{N}_n^I\})$ in which the
$N_n^I,\tilde{N}_n^I$ are not too large, when observed via probes of
energy ${\buildrel < \over \sim} \ \ {\alpha'}^{-1/2}$, will effectively
appear to be a particle with some internal structure of size $\sim
\sqrt{\alpha'}$
and momentum ${\bf p}$. (A large value of $N_n^I$ would elongate the size
in the $I^{th}$ direction to $\sim \sqrt{N_n^I\alpha'/n}$.
These are sometimes referred to as the `really stringy' states.)
It is natural to define the mass $M$ of a
state as  $M^2 = \epsilon^2 - {\bf p}^2$. Eq. (\ref{spectrum1})\ then
gives the mass formula in terms the oscillator excitations of the state. Thus
different states in the spectrum of a single string can be identified with
various particle species having different masses (which come in units of
$\alpha'^{-1/2}$) and momenta.

In addition to the length scale $\sa$, string theory has a dimensionless
coupling constant $g$, which represents the amplitude that two strings
touching each other will fuse into a single string, or the reverse
process. Between such joinings or splittings, strings travel freely
according to (\ref{action1}). These rules essentially specify string
perturbation theory completely. The effective interactions of
massless particles in the string spectrum (gravitons, dilatons and
antisymmetric tensor particles in the bosonic string and also photons
or gauge particles in the heterotic string) can be determined from
these considerations (see, e.g., Green, Schwarz and Witten 1987).
In particular the gravitational constant in $d$ spatial dimensions
is given by $G = g^2 {\alpha'}^{(d-1)/2}$. Thus Newton's constant
($G$ in $d=3$) is given by $G_N = g^2 \alpha'$ (assuming higher
dimensions compactify to radii $\sim \sa$). In other words, string
theory reproduces classical Einstein gravity at low energies if we
choose its two parameters $\alpha'$ and $g$ to satisfy
$g\sa = l_p$ (Planck length). Note that the `string length scale'
$\sa$ is $\sim l_p$ if $g \sim O(1)$ (strong coupling) and is much larger than
$l_p$ if $g \ll 1$ (weak coupling).

\vskip 0.3cm \noindent
{\bf The size of strings; string uncertainty principle}:
What is the size of the string in the state (\ref{wave1})? Consider the
transverse `mean-square spread' operator
(Mitchell and Turok 1987; Karliner, Klebanov and Susskind 1988)
$q \equiv \int_0^\pi d\sigma
({x^I(\sigma) - x^I_0})^2$. (\ref{wave1})\ is not an eigenstate of this,
but we can ask for its expectation value. Consider the ground state of
all the oscillator modes, $N_n^I=\tilde{N}_n^I=0$ (the scalar tachyon).
The expectation value of $q$ in this state
is $\langle q \rangle \sim \alpha'\sum_{n=1}^\infty 1/n$ which
diverges logarithmically because each of the
$x_n^I,\tilde{x}_n^I$ modes makes a finite contribution.
This divergence is empirically unobservable because an experiment does
not observe $q$ or the $x_n$ directly. A typical experiment involves
scattering a probe off the string. In order for the probe to `see' the
$x_n$ mode it must interact with it and excite it from its ground state.
This would cost energy $\sim \sqrt{n/\alpha'}$ from (\ref{spectrum1}).
A probe with a finite energy $E$
would only excite a finite number of oscillator modes; therefore the infinite
sum in $q$ should be cutoff at a finite value of $n$ depending upon the
energy of the probe. For $E \ll {\alpha'}^{-1/2}$, none of the oscillator
modes will be excited and the string will effectively look like a point
particle. Probes with $E \sim {\alpha'}^{-1/2}$ will see the state as a
fuzz of size $\sa$. For probes with energy
$E \gg {\alpha'}^{-1/2}$ the fuzz size will increase. The maximum fuzz
size is obtained if all the energy of the probe goes into exciting
only the $n=1$ mode. Its excitation level
is then $N_1^I \sim \alpha' E^2$ from (\ref{spectrum1}), and
the consequent root mean square spread in space of the target string
wavefunction (through its ${\bf x}_1$ and $\tilde{\bf x}_1$ modes)
is $\sqrt{N_1^I \alpha'}\sim \alpha' E$. The
size {\em grows} with the energy of the probe. This is a new term
that must be added to the usual uncertainty in position
$\Delta x \sim 1/E$ coming from Heisenberg's uncertainty principle.
Setting $\alpha' = G_N /g^2$ (in $3+1$ dimensions)
and putting back units we get the `string uncertainty principle'
\be
\label{stringup}
\Delta x \sim {{\hbar c} \over E} + {{G_N E} \over {g^2 c^4}}.
\ee
Minimizing this w.r.t. $E$, one finds the smallest observable
length scale in string theory ${\Delta x}_{min} \sim l_p/g \sim \sa$.
Here we assumed that all the energy goes into exciting only the $n=1$
modes. If the energy is shared with the higher modes whose
wavefunctions are more strongly localized, the spread will
be smaller. For example if one assumes that each mode $n$ upto
some maximum $n_{max}$ is excited to its first excited level
($N_n = 1$ for $n \leq n_{max}$ and zero thereafter), then one finds
$n_{max} \sim \sa E$ and the sum in $q$ should be cutoff at this
value. Then, instead of $\Delta x \sim \alpha' E$, one gets
$\Delta x \sim [\alpha' \ln (\sa E)]^{1/2}$, modifying the second
term in (\ref{stringup}).
Different choices putting in less energy into the higher modes
than the second case would
yield $\Delta x$ between these two values.
For all these choices it remains true that
${\Delta x}_{min} \sim \sa$.

These two forms of the string uncertainty principle were conjectured,
respectively, by
Gross (1989) and Amati, Ciafaloni and Veneziano (1989) from studies of string
scattering amplitudes at high energies at all loops
(Gross and Mende 1988; Amati, Ciafaloni and Veneziano 1988).
It is interesting that both forms as well as intermediate ones
can be derived by elementary considerations
of string wavefunctions using different assumptions of how the
energy is distributed among the oscillator modes of the target. One can
ask, what determines the actual distribution of the probe kinetic energy
among the various oscillator modes of the target? 
This needs further investigation.
The analysis of
Amati, Ciafaloni and Veneziano (1989) suggests that the scattering angle
plays a role in determining it.

{}From the above it is evident
that the smallest observable length of any object in a string universe
(where everything, objects and probes, is ultimately made of strings)
is of the order of $\sa$. This is a direct consequence
of the new wiggle degrees of freedom of strings.

\vskip 0.3cm
\noindent {\bf Gravitational collapse, black holes, random walks}:
In the above scattering experiment if too much
energy is deposited in the higher $n$ modes of the target string, its
size can become smaller than its Schwarzchild radius and it
can suffer gravitational collapse. For example,
an interesting choice of energy distribution among the modes of the
target is to assume that it is thermal. That is,
assume that all oscillator states of the target having a total
energy $E$ are equally likely. (To avoid a violation of unitarity,
the probe, also stringy, carries away all the correlations.)
What is $\langle q \rangle$ in such
an ensemble? This question has been investigated by Mitchell and Turok (1987)
and Aharonov, Englert and Orloff (1987)
in a different context. It turns out that $\langle q \rangle \sim
\alpha'^{3/2} E$. Taking $r = \langle q \rangle^{1/2}$ to be
the size of the string, such a string would suffer gravitational
collapse if its Schwarzchild radius exceeded $r$, or its energy
exceeded $\alpha'^{-1/2} g^{-4}$ (in $3+1$ dimensions). The entropy
$S$ of this `black hole', given by $\sim \sa E$ (see next section)
would be $S \sim \alpha' r^2$, proportional to its `area'.

The expectation value of the length of the string in the thermal
state is $\langle l \rangle \sim \alpha' E$, as for a classical
string configuration (or a cosmic string). Thus in this
state the string resembles a random walk in space with
step length $\sa$, since $l \sim \alpha'^{-1/2} r^2$ (Mitchell and
Turok 1987; Aharonov, Englert and Orloff 1987).

\vskip .3cm
\noindent {\bf String spectrum in a compact space}:
Spacetime itself is characterized by a metric (at least on scales
familiar to us), to determine which
we must measure lengths. If the smallest measureable length is in
principle $\sim \sqrt{\alpha'}$, this must ultimately reflect on the
smallest conceivable size of the universe in string theory. To study
this more precisely, we now consider strings in a finite sized space.

Consider a toroidal compactification of space with a radius $R$,
i.e., the coordinate $x^i$ of space ($i=1,\ldots, d$) is identified 
with $x^i + 2\pi w R$ with $w$ an integer. While we
describe this special case for simplicity and clarity, many of the
consequences for cosmology discussed later are valid for a much
larger class of compactifications. Then classical string configurations
have another mode, modifying (\ref{fourier}) to
\begin{equation}
\label{fourier1}
x^i(\sigma) = x^i_0 + 2L^i\sigma + \sum_{n=1}^\infty(x^i_n \cos 2n\sigma +
\tilde{x}^i_n \sin 2n\sigma),
\end{equation}
where $L^i = w^i R$ with $w^i$ an integer. As $\sigma$ runs from
$0$ to $\pi$, $x^i$ runs from $x^i(0)$ to $x^i(0) + 2\pi R w^i$, the
string therefore winds around the universe in the $i^{th}$ direction
$w^i$ times.

This adds a term $L^2/\alpha'$ to (\ref{lagrangian}), and $L^2/\alpha'^2$
to (\ref{spectrum1}). In compact space $p^i=m^i/R$ is also quantized
($m^i$ integer), and the spectrum is now given by (Green, Schwarz and Brink
1982)
\begin{equation}
\label{spectrum2}
\epsilon^2={{\bf m}^2 \over R^2} + {{{\bf w}^2 R^2} \over {\alpha'^2}}
+ (2/\alpha')[-2 + \sum_{I=1}^{d-1}\sum_{n=1}^\infty
n(N_n^I + \tilde{N}_n^I)].
\end{equation}
This spectrum maps into itself under the transformation
\be
\label{duality}
R \rightarrow \tR \equiv \alpha'/R.
\ee
This is evident from the fact that the state with quantum numbers
$({\bf m},{\bf w},\{N_n^I,\tilde{N}_n^I\})$
in a universe of radius $R$
has the same energy as the state
$({\bf w},{\bf m},\{N_n^I,\tilde{N}_n^I\})$
in a universe of radius $\tR$. The interchange ${\bf m} \leftrightarrow
{\bf w}$ together with the transformation (\ref{duality})\ does
not alter the r.h.s. of (\ref{spectrum2}). Thus at the level of the
free spectrum, string theory does not distinguish between a universe
of size $R$ and a universe of size $\tR$. This symmetry is also
respected by string interactions: the amplitude of a process
in a universe of size $R$
with a given set of external states
is the same as the amplitude in a universe of size $\tR$
of the `dual' set of states (obtained by
interchanging ${\bf m}$ and ${\bf w}$ quantum numbers for
each state in the first set). This symmetry, known as target-space
duality or `t-duality' was found by Kikkawa and Yamasaki (1984),
Sakai and Senda (1986), Nair, Shapere, Strominger and Wilczek (1987),
Sathiapalan (1987), and Ginsparg and Vafa (1987).

\vskip 0.3cm \noindent
{\bf A new periodic spatial coordinate and wavefunctions of winding states}:
In addition to the coordinate $x^i_0$ which is compact with
period $2\pi R$, there exists another spatial coordinate $\tilde{x}^i_0$
in string theory with period $2\pi \tR$ (Sathiapalan 1987).
This is just the conjugate variable to the operator
$\hat{L}^i \equiv (1/2\pi \alpha')\int_0^\pi d\sigma \ \partial_\sigma x^i$
whose eigenvalue is $L^i/\alpha'= w^i/\tR$ (just as $x^i_0$ is conjugate to
$\hat{P}^i$ whose eigenvalue is $p^i=m^i/R$). Formally, define
$|\tx\rangle \equiv \sum_{\bf w}e^{i\hat{\bf L} \cdot \tx}|{\bf w}\rangle$
and $\hat{\tilde{\bf x}}_0|\tx\rangle \equiv \tx|\tx\rangle$,
where the sum goes over ${\bf w} \in {\bf Z}^d$ and
$|{\bf w}\rangle$ denotes $|{\bf m},{\bf w},\{N_n^I,\tilde{N}_n^I\}
\rangle$ for brevity. It follows that
$[\hat{\tilde{x}}^i_0,\hat{L}^j] = i\delta^{ij}$.
Since ${\bf w}$ is quantized on an integer lattice, it is easy to see that
$|\tx + 2\pi {\bf n}\tR\rangle = |\tx\rangle$ for any
${\bf n} \in {\bf Z}^d$. I.e., the points $\tilde{x}^i_0$ and
$\tilde{x}^i_0 + 2\pi \tR$ in this `dual space' are physically
indistinguishable.
The wavefunctions are now given by
$\psi_{({\bf m},{\bf w},\{N_n^I,\tilde{N}_n^I\})}
({\bf x}_0,\tx,\{x_n^I,\tilde{x}_n^I\}) =
\langle {\bf x}_0,\tx,\{x_n^I,\tilde{x}_n^I\}|
{\bf m},{\bf w},\{N_n^I,\tilde{N}_n^I\}\rangle$, where the r.h.s differs from
that of (\ref{wave1}) by the factor $e^{-i{\bf p \cdot x_0}}$ being replaced
by $e^{-i({\bf m \cdot x_0}/R + {\bf w \cdot \tx}/\tR)}$. The physical
significance of this `dual position coordinate' will be discussed in
the last section.

\section{Statistical mechanics of strings at high energy densities}
\label{sec:statmech}
\noindent {\bf The partition function and the density of states}:
In order to study the very early universe in the context of string
theory, it is important to know how a very hot gas of superstrings
behaves. Consider the thermal partition function of a string gas:
\be
\label{Z}
Z(\beta,R)=\sum_{\alpha}\ e^{-\beta E_{\alpha}(R)}.
\ee
Here $\alpha = (N, a_1,\ldots,a_N)$ labels a state with $N$ strings,
the quantum numbers of the $k^{th}$ string being given by $a_k$.
Each $a_k$ in turn stands for the full set of quantum numbers
$({\bf m, w},\{N_n^I,\tilde{N}_n^I\})$ for the $k^{th}$ string. $E_{\alpha}(R)$
is the energy of the multi-string state $\alpha$ in a universe of
radius $R$, and in the ideal gas approximation is given by the
sum of the individual single string energies:
$E_{\alpha}(R) = \sum_{k=1}^N \epsilon_{a_k}(R)$
where $\epsilon_{a_k}(R)$ is given by (\ref{spectrum2}) for
closed bosonic strings. (For superstrings the formula for $\epsilon$
is modified by additional degrees of freedom but retains the
essential character needed for subsequent discussion.) The sum
over $\alpha$ includes a sum over all individual string states for
a fixed $N$ and a sum over $N$ from zero to infinity.

This partition function has a number of interesting properties.
First, it has singularities in the complex $\beta$ plane (other
than the usual $\beta=0$ singularity) even at finite volume. In
point particle field theories, singularities, which are usually signatures of
phase transitions, arise only in the thermodynamic limit. In the
string case they arise at finite volume because even a single
string has infinite degrees of freedom. The location of the right
most singularity,
$\beta_0$ ($\equiv 1/T_H$, where $T_H$ is known as the Hagedorn
temperature (Hagedorn 1965; Huang and Weinberg 1970)),
is proportional to the only length scale in the theory,
$\beta_0 =c_0\sqrt{\alpha'}$. The proportionality constant is
independent of the size of the box
(or universe) and other details of compactification
(Antoniadis, Ellis and Nanopoulos 1987; Axenides, Ellis and Kounnas 1988)
but dependent only
on the type of string theory, bosonic ($c_0=4\pi$), type II superstring
($c_0=2\sqrt{2}\pi$), or heterotic ($c_0=(2+\sqrt{2})\pi$).
As long as space is compact, the singularity is universally a simple
pole (Brandenberger and Vafa 1989; Deo, Jain and Tan (DJT) 1989a). There
is a representation of $Z(\beta)$ due to O'Brien and Tan (1987)
(see also Maclain and Roth 1987; McGuigan 1988)
which is useful in determining its analytic
structure in the complex $\beta$ plane. It turns out that there is
an infinite number of singularities to the left of $\beta_0$ (DJT 1989a)
whose locations in general depend upon the radius of universe.
For universes much larger than $\sqrt{\alpha'}$
(and also, by duality, for universes much smaller than $\sqrt{\alpha'}$),
a number of these singularities approach $\beta_0$.
Second, since the spectrum exhibits duality, so do the partition
function and density of states
$\Omega(E,R)= \sum_{\alpha}\delta (E-E_{\alpha}(R))$:
\be
\label{Zduality}
Z(\beta,R)=Z(\beta,\tilde{R})\quad \quad
{\rm and} \quad \quad \Omega(E,R)= \Omega(E,\tilde{R}).
\ee
This follows from the
fact that for every $\alpha$ there exists an $\tilde{\alpha}$
(obtained from $\alpha$ by interchanging the momentum and winding numbers
of every string in the state $\alpha$) such that
$E_{\alpha}(R)=E_{\tilde{\alpha}}(\tilde{R})$.
Third,
the behaviour of $Z(\beta)$ near $\beta_0$ is such that at temperatures
close to the Hagedorn temperature, fluctuations
are large and invalidate the use of the canonical ensemble for
deducing the thermodynamic properties of the string gas. One is forced
to use the more fundamental microcanonical ensemble, defined by $\Omega(E,R)$
(Frautschi 1971; Carlitz 1972; Mitchell and Turok 1987; Turok 1989).
Finally, since
$Z$ and $\Omega$ are related by a Laplace transform, the
leading large energy behaviour of $\Omega(E)$
is controlled by the behaviour of $Z(\beta)$ near its singularities,
and can be determined by a contour deformation technique
(DJT 1989a, 1991).
At large radius ($R \gg \sqrt{\alpha'}$) and at energy densities
above the `Hagedorn energy density' $\rho_0 \sim \alpha'^{-(\bar{d}+2)/2}$,
the density of states is given by (Deo, Jain, Narayan, Tan 1992 (DJNT))
\be
\label{density}
\Omega(E,R) \simeq \beta_0 \ e^{\beta_0 E + a_0 V}
[ 1-\delta(E,R)], \quad \quad
\delta(E,R)= {{(\beta_0 E)^{2\bar{d} - 1}} \over {(2\bar{d}-1)!}}
e^{-(\beta_0-\beta_1)(E-\rho_0 V)}.
\ee
Here we use the notation that $d$ represents the total number of
spatial dimensions, all of them compact ($d=25$ for bosonic strings
and $9$ for superstrings and heterotic strings). $\bar{d}$ is the
number of spatial dimensions that have large radius $R \gg \sqrt{\alpha'}$;
the remaining $d-\bar{d}$ dimensions are assumed to have radii $\sim
\sqrt{\alpha'}$. $V=(2\pi R)^{\bar{d}}$ is the volume of the large
dimensions. $a_0$ is a constant of order $\sim \alpha'^{-\bar{d}/2}$.
$\beta_1$ is the singularity of $Z(\beta,R)$ closest
to $\beta_0$; $\beta_0-\beta_1 \sim \alpha'^{3/2}/R^2$. The formula
(\ref{density})\ is valid for $\bar{d} > 2$ and for energy density
$\rho \equiv E/V$ greater than $\rho_0$. $\rho-\rho_0$ should be
large enough (greater than $O(\sqrt{\alpha'} R^{2-\bar{d}})$) so that
$\delta \ll 1$.

\vskip 0.3cm \noindent
{\bf Thermodynamic properties; physical interpretation in terms of
degrees of freedom}:
The thermodynamic properties of the gas are determined by (\ref{density}).
The entropy $S \equiv \ln \Omega$ is given by
\be
\label{entropy}
S(E,R) \simeq \beta_0 E + a_0 V + \ln (1-\delta),
\ee
from which one finds the temperature
$T \equiv [(\partial S/\partial E)_V]^{-1}$ to be
\be
\label{temp}
T(E,R) \simeq T_H(1-{{\beta_0-\beta_1} \over {\beta_0}}\delta),
\ee
and the pressure $p \equiv T(\partial S/\partial V)_E$
\be
\label{pressure}
p(E,R) \simeq T_H a_0 \big( 1-\delta{{\beta_0-\beta_1}\over \beta_0}
[1+{{\beta_0\rho_0}\over{a_0 \bar{d}}}
(2{\rho \over \rho_0} + \bar{d} - 2)]\big) .
\ee

Thus both the temperature and pressure of the string universe
reach asymptotic values determined by the string length scale
$\alpha'$ at energy densities above Hagedorn; corrections to
these asymptotic values are exponentially suppressed above these
energy densities. The physical reasons for this are as follows.
The leading contribution to the density of states of a string gas grows as
the exponential of a linear function of $E$,
unlike for a gas of point particles where it grows exponentially
with a sublinear function. This is because
the number of oscillator states at a fixed large value of
$\bar{N}\equiv \sum_{I=1}^{d-1}\sum_{n=1}^\infty
n(N_n^I + \tilde{N}_n^I)$, grows as
$\sim e^{c_1\sqrt{\bar{N}}}$.
This is just the Hardy-Ramanujam asymptotic formula for
the number of partitions of a large positive integer $\bar{N}$
into non-negative integers, a result in number theory. Thus, even
for a single string the density of states grows exponentially
$\sim e^{\beta_0 \epsilon}$ with energy (since $\sqrt{\bar{N}} \sim
\sa \epsilon$ from (\ref{spectrum2})). By contrast the contribution to
the density of states from the momentum and winding modes is
very small. E.g., for a single particle, for which only momentum modes
contribute, the density of states grows only as a power $\epsilon^{\bar{d}-1}$.
Thus at large energies it is entropically
favourable for the energy to go into oscillator modes rather than
momentum or winding modes. A term in the entropy of the gas that is
linear in energy
gives rise to a constant, i.e., energy independent temperature.
The form of the subleading corrections
(which is due to an interplay between oscillator, momentum and
winding modes) tells us that $T_H$ is
an upper limiting temperature.
Figure 1 displays the behaviour of temperature as
a function of energy for an ideal gas of strings and
contrasts it with an ideal gas of point particles.

The reason for the asymptotic pressure is as follows:
The leading term $\beta_0 E$ in the entropy
does not contribute to the pressure because it is independent
of the volume at constant energy; this is because the
oscillator mode contribution
to the energy (\ref{spectrum2})\ is volume independent.
The second term, $a_0 V$, should be
interpreted as the contribution of momentum modes.
It is proportional to volume
just like for a gas of point particles, due to the translational
degrees of freedom. (A pure winding mode gas
by contrast will give a contribution proportional to $1/V$,
which is subleading for large radii.)

For an ordinary point particle gas the entropy also depends upon
the energy density $S \simeq c_2 [\rho^{d/(d+1)}]V$, from which the
usual expression $p = \gamma \rho$ with
$\gamma = 1/d$ follows. In the string
case above, the coefficient of $V$ is just a constant, $a_0$.
The physical reason is that above Hagedorn energy density,
the energy density in momentum modes is a constant independent
of the total energy density. If more energy is pumped into
the box, it goes primarily into oscillator modes, which
are entropically favoured, than into momentum modes. Conversely, if
some energy is taken out of the box (keeping the total density still
above Hagedorn), it is primarily extracted from the oscillator modes
keeping the energy in the momentum modes essentially the same. Equivalently,
if one expands the volume slightly keeping energy the same
(this is what is implied by the derivative $(\partial/\partial V)_E$),
energy flows from the oscillators to the momentum modes to keep the
energy density in the latter constant. Thus the energy density in
momentum modes (which are the contributors to pressure, consisting
of small strings bouncing around like particles) is independent of
the volume or the total energy density (as long as the latter is
above Hagedorn) and hence is the pressure.

The above argument seems to be consistent with our present picture
of how the total energy of the gas is distributed among various
strings.
In the energy and radius domain under discussion, the
string gas can be considered to be consisting of broadly
speaking two `components', of energies $E_1$ and $E_2$
with $E=E_1+E_2$.
The first component consists of a few ($\sim \ln [R/\sa]$)
large strings which capture most of the energy of the gas
($E_1 \gg E_2$ provided $E \gg \rho_0 V$).
`Large' strings are those whose energies are $O(R^2 \alpha'^{-3/2})$
or greater. Most of their energy is due to oscillator modes and
the wavefunctions of these strings spread across the
whole universe (recall from the previous section that the size
of a thermal string of energy $\epsilon$ is $\sqrt{\alpha'^{3/2} \epsilon}$,
hence spread is $\sim R$ for $\epsilon \sim R^2 \alpha'^{-3/2}$).
If one adopts a classical picture, the universe is stuffed with
space filling brownian walks (see Salomonson and Skagerstam 1986;
Mitchell and Turok 1987). The second
component has fixed total energy $E_2 \sim \rho_0 V$ and
consists of many ($\sim V\alpha'^{-\bar{d}/2}$) small strings
`Small' ranges in size
from $O(\sqrt{\alpha'})$ to $< O(R)$, and in energy from
zero to $< O(R^2\alpha'^{-3/2}$). A crucial property of the
gas is that as more energy is pumped into
the box, it goes into the first component, leaving $E_2$ fixed.
This was qualitatively anticipated by Frautschi (1971), Carlitz (1972),
Mitchell and Turok (1987), Aharanov, Englert and Orloff (1987), and
Bowick and Giddings (1989), and made
quantitatively explicit in DJT (1989b, 1991) and DJNT. This picture is
unaltered
by the introduction of conservation laws for the total winding number
and momentum, even though additional subleading terms arise in the
density of states.

\vskip 0.3cm \noindent
{\bf Duality; thermodynamics in small spaces}:
We have so far discussed the case of large $E$ and large radius $R
\gg \sqrt{\alpha'}$. What happens at very small radii?
This is immediately answered by duality. At $R \ll \sqrt{\alpha'}$,
the r.h.s. of (\ref{density})\ has the same form but with $V$ replaced by
$\tilde{V} \equiv (2\pi\tilde{R})^{\bar{d}}$ (now $\beta_0-\beta_1
\sim \alpha'^{3/2}/{\tilde{R}^2}$). The same is therefore true of
temperature and pressure (we now define
$p \equiv T(\partial S/\partial \tilde{V})_E$).
At $R \sim \sqrt{\alpha'}$ we find
that the leading behaviour of the density of states is still given
by (\ref{density}), but now $V$ is replaced by a slowly varying
function of $R$ of order $\alpha'^{\bar{d}/2}$, and $\beta_0-\beta_1
\sim \sqrt{\alpha'}$ (DJT 1989a, DJNT). The
temperature as a function of $E$ is still given by (\ref{temp})\
with these replacements. However the pressure needs to be
appropriately defined and interpreted in this domain
(since $S(E,R)$ has to have an extremum at the duality radius,
both definitions $p \equiv T(\partial S/\partial V)_E$
and $p \equiv T(\partial S/\partial \tilde{V})_E$ imply that
$p$ passes through a zero at $R = \sa$).

\vskip 0.3cm \noindent
{\bf Inconsistency of string thermodynamics in non-compact spaces}:
Finally we remark that
string thermodynamics seems to be internally consistent
only in a compact space. The reason is that in a noncompact
space to define the density of states we have to consider an
artificial box of large volume $V$ to confine the gas and
later take the thermodynamic limit. This is problematic
in string theory because strings are extended objects,
they can in principle extend from one wall to another, and
render the entropy inextensive. One can see the problem
explicitly at high energy densities $E > \rho_0 V$ when
there exist a few strings in the gas whose individual
energy is a significant fraction of the total energy $E$.
The spread of their wavefunction is therefore $\sim \sqrt{\alpha'^{3/2} E}$.
Let the number of large dimensions (of radius $R \gg \sa$) be $\bar{d}$.
Then since $E > \alpha'^{-(\bar{d}+1)/2}R^{\bar{d}}$, these
strings have a size $\alpha'^{(2-\bar{d})/4} R^{\bar{d}/2}$.
Thus for $\bar{d} > 2$ these strings have a spread much greater
than $R$, the size of the universe itself. In a compact universe
this is not a problem; the string can wrap around the universe
many times. But if the universe were to be noncompact in these
$\bar{d}$ directions, then we find that these strings hit
the walls of the artificial box with nowhere to expand, leading
to an inconsistency of interpretation (see also DJNT).

\section{Implications for superstring cosmology and initial singularities}
\label{sec:cosmology}
\noindent {\bf Absence of a temperature singularity}:
We now discuss how the above considerations might impinge on cosmology.
Let us follow our present universe (assumed compact in all dimensions
but with three large dimensions)
backwards in time according to the standard model of cosmology.
At the epoch where the energy density in the large dimensions
is above $\rho_0 \sim \alpha'^{-2}$ but the radius is still much greater
than $\sa$
(this is quite natural in the standard model at early epochs),
let us assume that
the standard model physics is replaced by string theory,
and use the ideal gas approximation (\ref{density}). Then as
we proceed to smaller radii and hence higher energy densities, the temperature
and pressure being governed by equations (\ref{temp})\ and (\ref{pressure})
no longer increase indefinitely (as they would in any point particle
theory) but flatten out. The temperature remains flat as $R$ approaches
$\sqrt{\alpha'}$ and well into the domain $R \ll \sqrt{\alpha'}$
(as long as $E > \rho_0 \tilde{V}$ or $\tilde{\rho} \equiv E/\tilde{V} >
\rho_0$).
As $R$ declines further (i.e., $\tilde{R}$ increases) the temperature
{\em falls}. This is shown in figure 2. The behaviour of temperature
as a function of radius (at fixed energy or fixed entropy) is symmetric
about $R = \sqrt{\alpha'}$. At very small radius it does not diverge
as it does for a universe made of elementary point particles, but
behaves just as for a very large universe. The string universe has
no temperature singularity.

\vskip 0.3cm \noindent
{\bf Physical interpretation of a small universe}:
What is the physics of this bizarre behaviour? This was discussed
by Brandenberger and Vafa (1989), even before the precise expression
(\ref{density}) for the density of states was known. They asked the
question: how would one measure the size of the universe if it were
very small? For a large periodic box one can imagine sending a light signal
(a localized photon wave-packet) and measuring the time it takes to come back.
But this experiment would fail in very small box. The energy of a
momentum mode goes as $m/R$, and a superposition of many such modes
is needed to create a localized wave-packet, thereby making it more
and more energetically difficult to send a wave-packet in smaller universe.
In string theory the photon is a massless state with some momentum quantum
number $\bf m$, winding number zero, and a single oscillator excitation
(the term $[-2 + \bar{N}]$ in (\ref{spectrum2}), or its analogue for
heterotic strings, is zero). In today's universe (assumed large) these 
are easily excited, but it would be energetically very difficult
to create photons and send them around in a universe of size
$R \ll \sqrt{\alpha'}$ (see (\ref{spectrum2})). On the other hand, in
a very small universe, particles `dual' to the photon, with quantum
numbers ${\bf m}=0$, some winding number {\bf w}, and the same oscillator
quantum numbers as the photon would be easily excited. Indeed these
would constitute the `light' particles of the very small universe. An
observer in this very small universe would hardly think of sending
photons to measure the size of his universe (just as we would not
contemplate sending winding modes around); he would use a superposition
of the `dual photon' modes. By sending such modes he would be measuring
the extent of the `dual position coordinate' $\tilde{x}^i$ (recall that
$\tilde{x}^i$ is to winding
modes what position $x^i$ is to momentum modes). But, as discussed
earlier, that extent is just ${2\pi \tR}$; hence observers in a universe
of size $R \ll \sqrt{\alpha'}$ would find its radius to be not $R$ but
$\tilde{R} = \alpha'/R \gg \sqrt{\alpha'}$.

Indeed in a universe with $R \ll \sqrt{\alpha'}$ all momentum modes
would be energetically difficult to excite. Everything - signals,
apparatus, observer - would be made from particles that have zero
{\bf m} quantum number (in our present large universe everything is made
of zero {\bf w} quantum number). Since string theory has
duality as a symmetry of the spectrum as well as the interactions,
the dual particles would interact with each other exactly the way
normal particles do in our present universe. The observers in a very
small universe would not therefore know that they are in a universe
much smaller than $\sqrt{\alpha'}$, their physics would be identical
to ours (and for that matter nor do we know whether our universe
is very large or very small compared to $\sa$).

It is therefore no surprise that temperature has the behaviour shown
in figure 2. As radius goes much smaller than $\sqrt{\alpha'}$,
the universe actually {\em expands, as seen by the modes
that are excited
in it}. This also makes it evident that there are no physical singularities
in the energy density, pressure or curvature as $R \rightarrow 0$. In a very
small universe, the physical energy density is not $E/V$ but
$E/\tilde{V}$ (which goes to zero and not infinity as $R \rightarrow 0$),
since the physical volume of the universe is $\tilde{V}$.
In string theory the smallest {\em physical} size of the universe
is $\sa$.

Note that the arguments leading to the string uncertainty
principle - that the smallest observable size of an elementary string is
$\sa$ - and the arguments leading to the same minimum physical size of the
universe both make essential use of probes in thought experiments. Also
note the difference: while the former argument uses
the oscillator modes, the latter rests on the duality between
momentum and winding modes (although the limiting temperature and pressure
depend again on the oscillators). All these
modes are simultaneously forced upon us as soon as we accept strings as the
elementary constituents of nature, and all are governed by a single scale
parameter that appears in (\ref{action1}).

\vskip 0.3cm \noindent
{\bf A cosmological scenario without initial singularities}:
Brandenberger and Vafa sketch the following scenario.
Let us assume that
at some point in the future our universe stops expanding and starts
contracting and heating up. As the energy density increases to the Hagedorn
energy density, stringy effects will take over and the temperature
will flatten out. If it continues to contract through the duality
radius and
comes out the `other side', then dual (analogues of winding) modes
will take over.  The universe will cool
and `expand' and give rise to dual nucleons, galaxies, stars, planets, life,
etc.
What appears to us to be the `big crunch' will
be a `big bang' for the dual observers. The process could repeat
giving rise to an oscillatory universe. `Our own' big bang was just one
such periodic occurrence.

Of course much more work is needed to justify any such {\em dynamical}
scenario. We have been concerned with just those aspects which hinge
only on the {\em degrees of freedom}. A body of literature
now exists which also deals with the time evolution of the metric
and other low energy modes in string theory
in the cosmological context (see Tseytlin and Vafa (1992), Gasperini (1997),
the contribution by Bose (1997) to these proceedings, and references therein).
Perhaps it would be worthwhile to revisit some of this
in the light of the expression for the pressure of a string gas
presented here, since pressure as part of the energy momentum tensor
is a source in the field equations.\footnote{This suggestion arose in
discussions with S. Kalyana Rama.}

Nevertheless the above scenario is important in that it at least
allows us to {\em imagine} how initial singularities might be avoided in
string theory.
It is important to emphasize that singularities are avoided not by
recourse to quantum gravity (spacetime has all along been treated
classically) but simply by a reinterpretation of what it means to
talk of a small universe in the light of string theory.
In point particle theories, classical imagination fails at $R=0$.
This is an example of how the new
degrees of freedom in string theory allow
(or rather, necessitate) a new perspective on our ideas of spacetime,
in this case specifically on our notion of the size of the universe.
It should be mentioned that while we have explicitly discussed
the case of a toroidal compactification for simplicity, the t-duality
symmetry which makes this reinterpretation possible holds for
a much larger class of string models (and is expected to be a
symmetry in a nonperturbative formulation of string theory).
A limiting temperature and pressure in the ideal gas approximation
also seem to be a universal feature of strings in compact
spaces.

\vskip 0.3cm \noindent {\bf Cautionary remarks}:
At this point some caveats are in order. Thermodynamics
in the presence of gravity must take into account the
Jeans instability. At constant energy density a sufficiently
large volume will be susceptible to gravitational collapse. This
places an upper limit on the value of $R$ for which our thermodynamic
considerations are valid. Second, the results are
based on an ideal gas approximation, used in a regime of
high energy densities, greater than the string energy scale itself.
This is justified only if the coupling is weak ($g \ll 1$).
Even for a fixed weak coupling the approximation can be expected to break
down at sufficiently high energy densities, at which point
non-perturbative effects will need to be taken into account.
This places a lower limit on $R$ for the validity of the
approximation. Thus there is possibly a window $R \in (R_1,R_2)$,
$\sa < R_1 \leq R_2$ (and the `dual window' $\tR \in (R_1,R_2)$)
in which one can expect this to be valid. The
window expands in both directions as coupling becomes weaker
(see Atick and Witten (1988) for related arguments).

In addition to string interaction and nonperturbative effects in the
region of $R$ close to $\sa$, we face the uncertainty of interpretation
of spacetime itself at such small scales. For sufficiently large
(or sufficiently small) $R$, spacetime may be treated classically,
as we have done. But this is questionable
near the duality radius. This is the regime where the universe as
well as its elementary constituents have the same `size'.
This problem awaits a better understanding
of spacetime in string theory.

\vskip 0.3cm \noindent
{\bf Possible observational consequences, further questions}:
Assuming that there was an era in the past where
the ideal string gas approximation was valid, could there be
some observable relic? From the picture of how energy is distributed
in the string gas it seems likely that density fluctuations
would have a different character in the stringy era, and as seeds
for later structure formation could have observable consequences.
Second, it would be interesting to look for signatures of compactness at
very large scales in the universe.\footnote{I thank T. Souradeep
for informing me that such analyses of the data are possible.}
Apart from a resolution of the singularity problem,
string thermodynamics also seems to be internally consistent only
in compact spaces. A compact universe is even otherwise natural in string
theory, since the extra dimensions in any case have to be compact.

At a more theoretical level, it may be worthwhile to investigate
dynamical mechanisms based on string modes (see Brandenberger and
Vafa (1989) for a proposal) for why only three spatial dimensions
are large. Also it is of interest to study how the recent progress
in our understanding of some non-perturbative aspects of string theory
affects the above considerations.

Note added: After this writeup was submitted for the proceedings, I
became aware of other papers (Yoneya 1989; Konishi, Paffuti and
Provero 1990; Kato 1990; Susskind 1994) which attempt to derive the
string uncertainty principle. The argument presented in the 
present article is different from those given in these papers.
Other recent references of related interest are Li and Yoneya 1997,
which argues that the string uncertainty principle is consistent
with D-brane dynamics, as well as Barbon and Vazquez-Mozo 1997 and
Lee and Thorlacius 1997, which attempt to include D-branes within
superstring statistical mechanics. For literature on a minimal length
in the context of quantum gravity without invoking string theory see
references in the review by Garay 1995, as well as Padmanabhan 1997.

I thank N. Deo, C-I Tan and C. Vafa for discussions in which most
of my understanding of string thermodynamics and cosmology was
developed, S. Kalyana Rama for getting me interested in the pressure
of an ideal string gas and pointing out some recent references, 
and the participants and organizers of
the Conference on Big Bang and Alternative Cosmologies for a
stimulating meeting.

\begin{figure}[htb]
\epsffile[100 390 650 750]{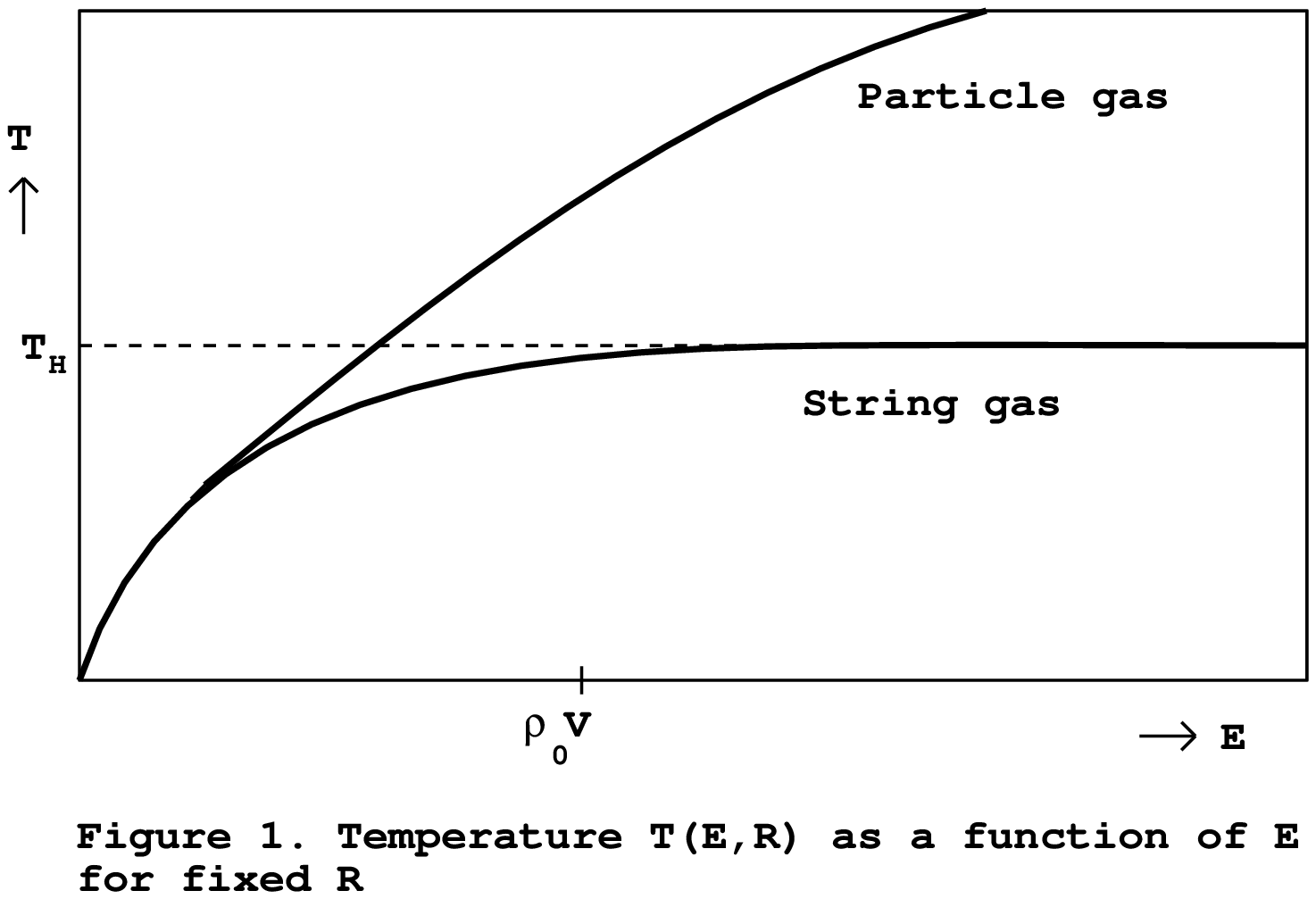}
\end{figure}
\begin{figure}[htb]
\epsffile[100 390 650 750]{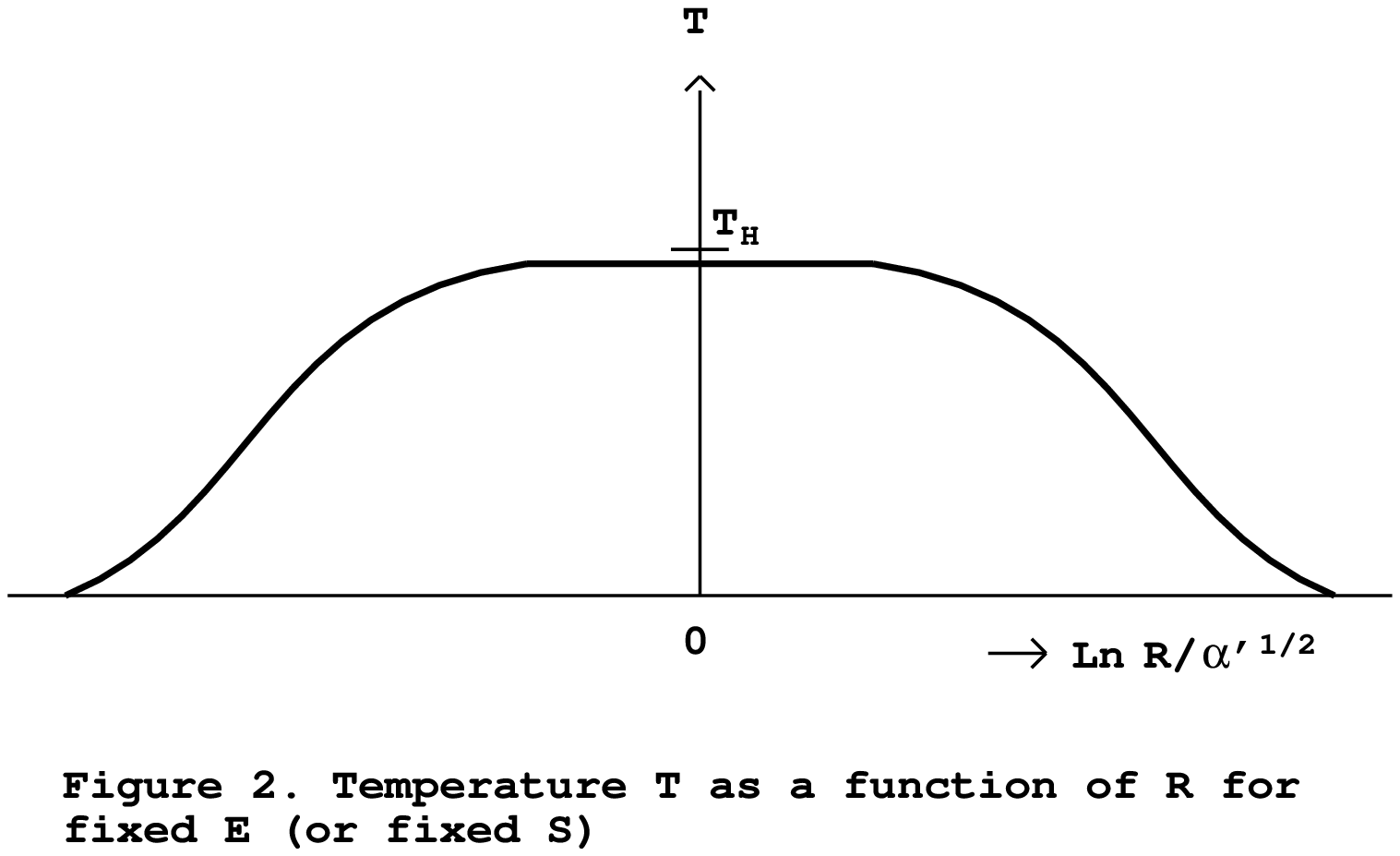}
\end{figure}

\end{document}